\documentclass[amsmath,amssymb,aps,prd,nofootinbib,reprint,superscriptaddress]{revtex4-1}

\usepackage{aas_macros,bm,CJK,graphicx,hyperref,soul,xcolor}
\usepackage[capitalize,noabbrev]{cleveref}

\newcommand{\bu}{{\bm u}}
\newcommand{\br}{{\bm r}}

\definecolor{tabblue}{HTML}{1f77b4}
\definecolor{taborange}{HTML}{ff7f0e}
\definecolor{tabgreen}{HTML}{2ca02c}
\definecolor{tabred}{HTML}{d62728}
\definecolor{tabpurple}{HTML}{9467bd}

\newtheorem{definition}{Definition}

\begin{document}

\title{Photon Ring Interferometric Signatures Beyond The Universal Regime}

\author{He Jia (\begin{CJK*}{UTF8}{gbsn}贾赫\end{CJK*})}
\email{hejia@princeton.edu}
\affiliation{Department of Astrophysical Sciences, Princeton University, Princeton, NJ 08544, USA}

\author{Eliot Quataert}
\affiliation{Department of Astrophysical Sciences, Princeton University, Princeton, NJ 08544, USA}

\author{Alexandru Lupsasca}
\affiliation{Department of Physics \& Astronomy, Vanderbilt University, Nashville, TN 37212, USA}

\author{George N. Wong}
\affiliation{School of Natural Sciences, Institute for Advanced Study, 1 Einstein Drive, Princeton, NJ 08540, USA}
\affiliation{Princeton Gravity Initiative, Princeton University, Princeton, NJ 08544, USA}

\date{\today}

\begin{abstract}
We calculate the interferometric signatures of black hole photon rings beyond the universal regime by perturbatively including the effects of finite ring width.
Our approach first slices a thick ring into a series of thin rings, each of which falls within the universal regime.
We thus calculate the visibility of the thick ring by aggregating the contributions from each thin ring, and then perturbatively expand the result into polynomials of the baseline length $u$.
We show that the visibility amplitude of a thick ring depends on its ``center-of-light'' diameter; it also includes additional higher-order corrections due to the width of the ring, with the leading correction terms proportional to $u^2$ for the envelope and $u^3$ for the phase.
We apply our method to images ray traced from general-relativistic magnetohydrodynamic (GRMHD) simulations and demonstrate that incorporating the higher-order corrections is crucial for accurately modeling the visibility of the first photon ring around M87*.
\end{abstract}

\maketitle

\section{Introduction}

The first, 230\,GHz, Event Horizon Telescope (EHT) images of the supermassive black holes (SMBH) M87* \cite{eht2019m87i} and Sgr~A* \cite{eht2022sgri} feature a bright, asymmetric ring, in complete accordance with ray traced images based on general-relativistic magnetohydrodynamic (GRMHD) simulations \cite{eht2019m87v}.
Theoretical models predict that, after enough time-averaging, the small-scale structure in such SMBH images comes to be dominated by a series of (not yet observed) narrow lensed photon rings \cite{johnson2020universal,gralla2020lensing,gralla2020shape,cardenas2023adaptive} exhibiting a \textit{universal} interferometric signature from which one can infer the projected diameter of the ring.
This observable diameter encodes information about the black hole mass and spin \cite{johnson2020universal,paugnat2022photon}, and can be measured using  very-long-baseline interferometric (VLBI) observations on a \textit{single} baseline extending from a ground-based telescope to one orbiting satellite, without requiring a dense sampling of the visibility plane to reconstruct the full image.

Prior studies \cite{johnson2020universal,gralla2020measuring,gralla2020diameter} of the interferometric signature of a thin ring with diameter $d$ and width $w\ll d$ have shown that in the \textit{universal} regime $1/d \ll u \ll 1/w$, the visibility amplitude displays a ringing pattern of weakly damped oscillations with respect to the baseline length $u$.\footnote{In this paper, a ring is considered ``thin'' if $w\ll1/u$, and ``thick'' if $w\gtrsim1/u$.
Therefore, the ``thickness'' of the ring depends on the baseline length $u$ of the VLBI observation.}
While the envelope of the oscillations depends on the detailed intensity profile of the ring, their frequency only depends on the ring diameter in the zero width limit, thus enabling a measurement of the ring diameter from the visibility amplitude alone.
Unfortunately, since realistic photon rings have finite width, the condition $u\ll1/w$ may not hold in practice, especially for the $n=1$ photon ring, which will be the first target of future high-resolution space-VLBI observations.

In this work, we study the interferometric signatures of finite width photon rings via a perturbative approach.
We show that, beyond the universal regime, the visibility amplitude of a finite-width, \textit{non-radially-symmetric} ring displays a baseline-dependent oscillation frequency (the precise meaning of this will be clarified in what follows).
In principle, this implies that observations on different baselines measure different ring diameters (due to finite ring-width effects), leading to some bias in the diameter inference when analyzing VLBI data (and hence, a bias in the inference of physical parameters from such data).
We show that this bias can be modeled by including in the visibility phase correction terms proportional to $u^3$, $u^5$, etc.
Our results also clarify what precise diameter is measured by VLBI observations of a finite-width ring: the \textit{center-of-light} diameter defined in Sec.~\ref{sec:ThickRings} below.

The remainder of this paper is organized as follows.
In Sec.~\ref{sec:ThinRings}, we review the visibility calculations of Ref.~\cite{gralla2020measuring} for general convex thin rings within the universal regime.
Next, in Sec.~\ref{sec:ThickRings}, we introduce the notion of \textit{sliceable} thick rings and describe an algorithm for decomposing a finite-width ring into a sum of thin rings.
We then calculate the visibility of a sliceable ring in a perturbative expansion in $uw\lesssim1$, which allows us to push beyond the universal regime.
We apply our method to ray traced GRMHD images in Sec.~\ref{sec:Numerics}, and we find that keeping the higher-order corrections is necessary for accurately modeling the visibility for M87*.
Finally, we conclude in Sec.~\ref{sec:Discussion} with a discussion of the implications of our results for spaceborne measurements of the photon ring, as well as potential directions for future research.
We provide a \texttt{python} implementation of our method to slice a thick ring and calculate its visibility in the publicly available package \texttt{hiring}.\footnote{\url{https://github.com/h3jia/hiring}.}

\section{Thin ring visibility revisited}
\label{sec:ThinRings}

In this section, we briefly review the prior work \cite{gralla2020measuring} on the visibility of a thin, convex ring that forms the basis of our subsequent study of finite-width rings.
In Sec.~\ref{sec:ThinConvexRing}, we follow Gralla's analysis \cite{gralla2020measuring} to obtain the visibility of a general thin ring in the large-$u$ limit, assuming a smooth angular profile for the ring.
Alternatively, in Sec.~\ref{sec:ThinFourierRing}, we derive the exact visibility of a \textit{circular}, non-axisymmetric thin ring, before taking the same large-$u$ limit to show that the result agrees with that of Sec.~\ref{sec:ThinConvexRing}.

\subsection{General derivation for convex rings}
\label{sec:ThinConvexRing}

An interferometer samples the radio visibility (the complex Fourier transform) of the sky intensity $I(x,y)$,\footnote{Throughout this paper, we use the symbol $I$ to denote the \textit{surface} intensity, such that the total flux density is $\int I(x,y)\,{\rm d}x\,{\rm d}y$, and $\mathcal{I}$ for the \textit{line} intensity, such that the total flux density is $\int \mathcal{I}(s)\,{\rm d}s$.}
\begin{align}
	V(\bu)=\int I(\br)\,e^{-2\pi i\bu\cdot\br}\,{\rm d}^2\br,
\end{align}
where $\br=(x,y)$ is the sky position in radians while the baseline $\bu=(u_x,u_y)$ $=(u\cos\varphi_{\rm obs},u\sin\varphi_{\rm obs})$ is the distance (in observation wavelengths) between telescopes projected into the plane perpendicular to the line of sight.
Without loss of generality, we can assume that $\bu$ lies along the $x$-axis (i.e., that $\varphi_{\rm obs}=0$), so that $V(\bu)$ only depends on the projected intensity $\mathcal{P}_x I(x)=\int I(\br)\,{\rm d}y$,
\begin{align}
    V(u)=\int\mathcal{P}_xI(x)\,e^{-2\pi i u x}\,{\rm d}x.
\end{align}
For an infinitely thin ring parameterized by its arclength $s$ as $(x_s(s),y_s(s))$, the sky intensity is
\begin{align}
    I(x,y)=\int\mathcal{I}(s)\,\delta(x-x_s(s))\,\delta(y-y_s(s))\,{\rm d}s,
\end{align}
and thus, the projected intensity is given by
\begin{align}
	\label{eq:Projection}
    \mathcal{P}_xI(x)=\int I(x,y)\,{\rm d}y
    =\sum_{\rm branches}\frac{\mathcal{I}(s)}{{\rm d}x_s/{\rm d}s},
\end{align}
where the sum runs over all $s$ such that $x=x_s(s)$.
If for some $s_0$ in the sum, $x_s'(s_0)=0$ with $\mathcal{I}(s_0)\neq0$, then $\mathcal{P}_xI(x)$ diverges at $x_0=x_s(s_0)$, where it behaves as $\mathcal{I}(s)/|{\rm d}x_s/{\rm d}s|\approx\mathcal{I}(s_0)/\sqrt{|2(x-x_0)/\mathcal{R}_0|}$, with $\mathcal{R}_0$ the curvature radius of the ring at $x_0$ \cite{gralla2020measuring,gralla2020diameter}.

Heuristically, the visibility at large $u$ is dominated by the small-scale features in $\mathcal{P}_xI$.
Assuming that $\mathcal{I}(s)$ is everywhere positive with characteristic correlation scale $\xi_\mathcal{I}$, the visibility at $u\gg\max(1/d,1/\xi_\mathcal{I})$ is dominated by the $|x-x_0|^{-1/2}$ divergences in $\mathcal{P}_xI$ introduced by the projection effects at the edge points where ${\rm d}x_s/{\rm d}s=0$.

Strictly speaking, $\xi_\mathcal{I}$ should also account for any ``jaggedness'' in the ring shape $(x_s(s),y_s(s))$, which can also introduce small-scale structure in the projection \eqref{eq:Projection} beyond the $|x-x_0|^{-1/2}$ singularities near the edge points, even if $\mathcal{I}(s)$ is smooth.
To our knowledge, the condition $u\gg1/\xi_\mathcal{I}$ has not been explicitly noted before.

For a convex shape, there are only two edge points, namely the leftmost ($x=x_L$) and rightmost ($x=x_R$) points in the projected intensity profile, which together define the projected diameter $d=x_R-x_L$ of the ring.
The Heaviside step function $H(x)$ has Fourier transform
\begin{align}
	\label{eq:HeavisideFT}
    \int\frac{H(x)}{\sqrt{x}}\,e^{-2\pi iux}\,{\rm d}x=\frac{e^{-\frac{i\pi}{4}{\rm sign}(u)}}{\sqrt{2|u|}},
\end{align}
which lets us compute  the visibility of an infinitely thin, convex ring in the limit of large $u\gg\max(1/d,1/\xi_\mathcal{I})$:
\begin{align}
    \label{eq:UniversalVisibility}
    V(u)=A_Le^{-\frac{i\pi}{4}}\frac{e^{-2\pi ix_Lu}}{\sqrt{u}}+A_Re^{\frac{i\pi}{4}}\frac{e^{-2\pi ix_Ru}}{\sqrt{u}}.
\end{align}
where $A_i=\mathcal{I}(x_i)\sqrt{\mathcal{R}_i}$.
Having obtained this formula, we conclude that it continues to hold for a thin ring (of width $w\ll 1/u$), which must produce the same interferometric response on baselines too short to resolve its profile.

In practice, because of the difficulties inherent in VLBI phase calibration, it is likely that in the near future we will only be able to measure the visibility amplitude
\begin{align}
	\label{eq:UniversalAmplitude}
    |V(u)|^2=\frac{1}{u}\left[A_L^2+A_R^2+2A_LA_R\sin(2\pi du)\right], 
\end{align}
which gives the \textit{universal} interferometric signature of a (thin) black hole photon ring.
From Eq.~\eqref{eq:UniversalAmplitude}, it is evident that the oscillating frequency of $|V(u)|^2$ with respect to $u$ is determined by the projected diameter $d$, while the angular asymmetry ($A_L\neq A_R$) of the ring only affects the amplitude of the oscillation.

To summarize, Eq.~\eqref{eq:UniversalAmplitude} gives the visibility of a (convex) thin ring in the \textit{universal} regime
\begin{align}
	\label{eq:UniversalRegime}
    \max(1/d,1/\xi_\mathcal{I})\ll u\ll1/w,\quad
    \mathcal{I}_L\mathcal{I}_R>0.
\end{align}
In this regime, the visibility amplitude takes the universal form \eqref{eq:UniversalAmplitude}, such that the locations of its peaks and valleys depend only on the projected diameter $d$ of the ring.

Finally, when $\mathcal{I}_L\mathcal{I}_R=0$, at least one of the left/right singularities in the projected intensity disappears, and hence Eq.~\eqref{eq:UniversalAmplitude} may not accurately represent the visibility (which may instead be dominated by other features in the intensity profile).\footnote{Mathematically, if $\mathcal{I}_L\mathcal{I}_R<0$, then the oscillation undergoes a $180^\circ$ phase shift, so that the peaks become valleys and vice versa.
Nevertheless, physical photon rings must have $\mathcal{I}>0$ everywhere.}
In the next section, we show explicit examples where the condition $u\gg1/\xi_\mathcal{I}$ fails, so that the locations of the peaks and valleys shift as a result.

\subsection{Fourier-based derivation for circular rings}
\label{sec:ThinFourierRing}

\begin{figure}[h!]
	\centering
	\includegraphics[width=\columnwidth]{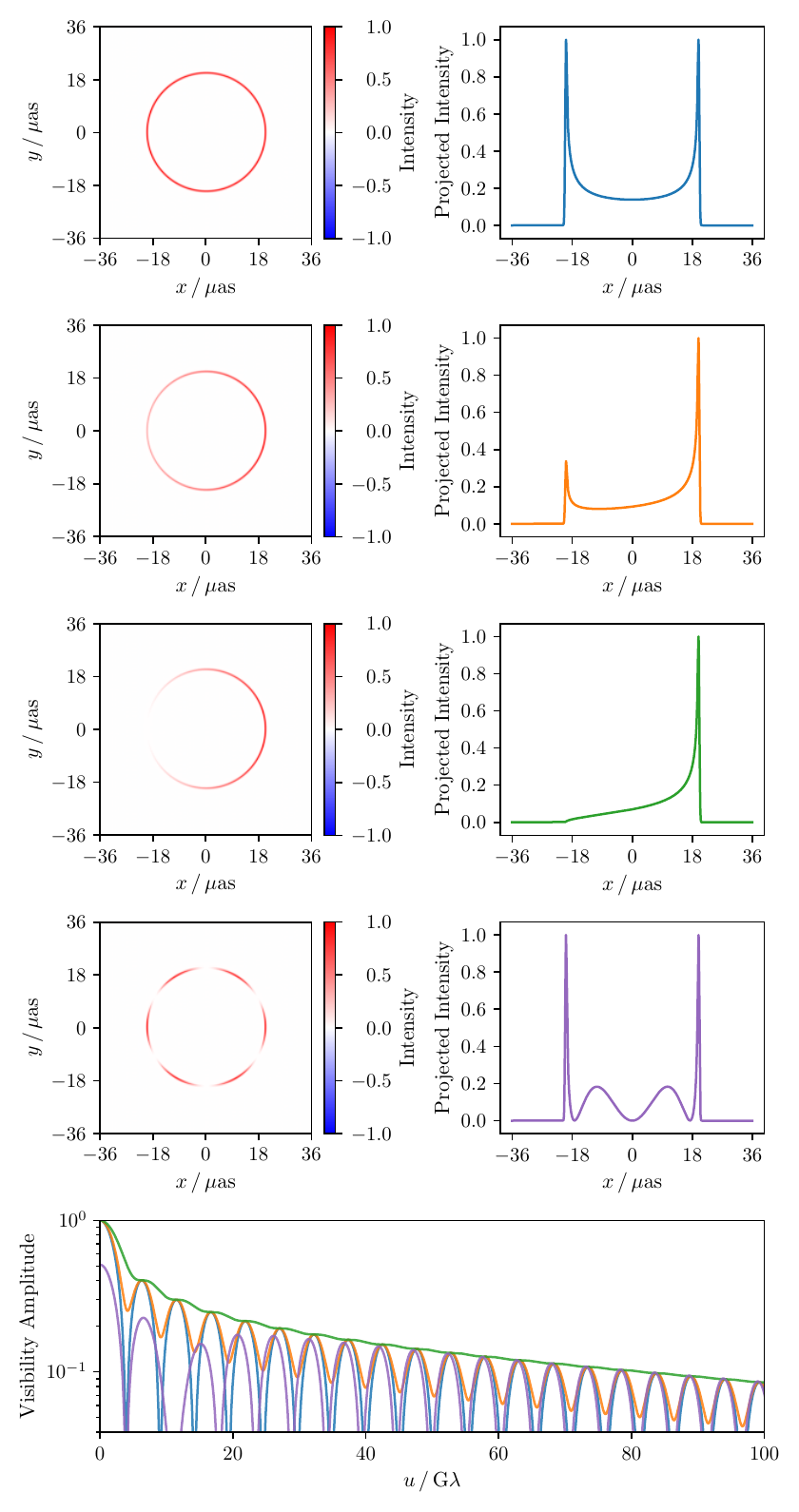}
	\caption{From top to bottom, we show the images of thin rings with $\mathcal{I}(\phi)$ equal to \textcolor{tabblue}{$\mathcal{I}_0$}, \textcolor{taborange}{$(2+\cos{\phi})\,\mathcal{I}_0/3$}, \textcolor{tabgreen}{$(1+\cos{\phi})\,\mathcal{I}_0/2$}, and \textcolor{tabpurple}{$(1+\cos{6\phi})\,\mathcal{I}_0/2$}.
	In the very bottom panel, we compare the visibility amplitude of the different rings as a function of $u$, after slightly adjusting the normalization of each ring for a clearer comparison.
	Angular asymmetry only changes the peak-to-valley ratio of the oscillations.
	However, if small-scale variation is present in $\mathcal{I}(\phi)$ (as in the \textcolor{tabpurple}{purple} model), then the peaks and valleys in $|V(u)|$ only align with those predicted by the universal form \eqref{eq:UniversalAmplitude} on the longer baselines $u\gg1/\xi_\mathcal{I}$, as expected from the conditions \eqref{eq:UniversalRegime}.
	The visibility amplitude of the \textcolor{tabgreen}{green} model does not display clear oscillations, as $\mathcal{I}_L=0$ and hence their amplitude vanishes according to Eq.~\eqref{eq:FourierVisibility}.}
	\label{fig:ThinRings}
\end{figure}

In the last derivation, we first argued---following Gralla \cite{gralla2020measuring}---that in the large-$u$ limit, the visibility is dominated by the $|x-x_0|^{-1/2}$ divergences in the projection of the ring intensity, and then we approximated the visibility of the ring as the sum of the (shifted) visibilities of the leftmost and rightmost singularities.
In this section, we first derive the exact visibility (valid for any $u\ll1/w$) of a \textit{circular} yet non-axisymmetric thin ring, and then we demonstrate that in the $u \gg \max(1/d,1/\xi_\mathcal{I})$ limit, our result is consistent with Eq.~\eqref{eq:UniversalAmplitude}.

First, a general circular thin ring may be decomposed into Fourier modes $\beta_m$ as
\begin{align}
    I(\rho,\phi)&=\delta\left(\rho-\frac{d}{2}\right)\,\mathcal{I}(\phi),\quad
    \mathcal{I}(\phi)=\frac{1}{\pi d}\sum_{m=-\infty}^{\infty}\beta_m e^{im\phi},\notag
\end{align}
and its visibility is then given by \cite{johnson2020universal}
\begin{align}
    \label{eq:BesselExpansion}
    V(u)=\sum_{m=-\infty}^\infty(-i)^m\beta_mJ_m(\pi du),
\end{align}
where $J_m(\cdot)$ denotes the $m^\text{th}$ Bessel function.
Again, we assume that $\bu$ is parallel to the $x$-axis; for the projection along a different direction, one can simply rotate the $\beta_m$ coefficients and our main conclusions here will still hold.
As shown in App.~\ref{app:CircularThinRing}, in the long-baseline limit, we have
\begin{align}
    &V(u)=\frac{\mathcal{C}_{\rm e}\cos(\pi du-\frac{\pi}{4})+i\,\mathcal{C}_{\rm o}\sin(\pi du-\frac{\pi}{4})}{\sqrt{u}},\notag\\
    \label{eq:FourierVisibility}
    &|V(u)|^2=\frac{\left(\mathcal{C}_{\rm e}^2+\mathcal{C}_{\rm o}^2\right)+\left(\mathcal{C}_{\rm e}^2-\mathcal{C}_{\rm o}^2\right)\sin(2\pi du)}{2u},
\end{align}
where the even and odd coefficients $\mathcal{C}_{\rm e}$ and $\mathcal{C}_{\rm o}$ are real, determined by the $\beta_m$, and satisfy
\begin{align}
    \label{eq:EvenOdd}
    \mathcal{C}_{\rm e}+\mathcal{C}_{\rm o}=\sqrt{2d}\,\mathcal{I}_{L},\quad
    \mathcal{C}_{\rm e}-\mathcal{C}_{\rm o}=\sqrt{2d}\,\mathcal{I}_{R},
\end{align}
such that Eq.~\eqref{eq:FourierVisibility} is consistent with Eq.~\eqref{eq:UniversalAmplitude}.

For a physical photon ring, $\mathcal{I}_L>0$ and $\mathcal{I}_R>0$, so that $\mathcal{C}_{\rm e}^2-\mathcal{C}_{\rm o}^2>0$.
Hence, the peaks and valleys of $|V(u)|$ are always in the same location for any ring of diameter $d$, regardless of its angular profile.
The approximation \eqref{eq:FourierVisibility} follows from the large-argument expansion of $J_m(\pi d u)$, which is valid only if $\pi du\gg m$ \cite{courant2008methods}.
This amounts to the condition $u\gg1/\xi_\mathcal{I}$, since $\xi_\mathcal{I} \approx d/m$ for the $m^\text{th}$ Fourier component: if $\mathcal{I}(\phi)$ has substantial contribution from large-$m$ modes, then the universal interferometric signature of the ring must be sampled on longer baselines $u$ in order for the power due to $\xi_\mathcal{I}$ to become negligible.

In Fig.~\ref{fig:ThinRings}, we compare some toy models of circular thin rings and their visibilities.
In the universal regime $\max(1/d,1/\xi_\mathcal{I})\ll u\ll1/w$, the visibility amplitude is given by Eqs.~\eqref{eq:UniversalAmplitude} or \eqref{eq:FourierVisibility}, and the peaks and valleys align for all rings.
For $\mathcal{I}(\phi)=(1+\cos{6\phi})\,\mathcal{I}_0/2$, the condition $u\gg1/\xi_\mathcal{I}$ is not met for $u\lesssim50\,\text{G}\lambda$, as can be seen in the phase difference between the purple and blue/orange curves in the bottom panel of Fig.~\ref{fig:ThinRings}.
For $\mathcal{I}(\phi)=(1+\cos{\phi})\,\mathcal{I}_0/2$, the oscillation in the visibility amplitude vanishes altogether because $\mathcal{I}_L=\mathcal{C}_{\rm e}^2-\mathcal{C}_{\rm o}^2=0$.

\section{Effects of finite ring width}
\label{sec:ThickRings}

We now study the visibility of rings of finite width $w$ in the region where the condition $u\ll1/w$ no longer holds.
We find that the ring width may introduce additional $u$-dependent corrections to the oscillating phase of the visibility amplitude, which must be correctly modeled to ensure an unbiased estimate of the ring diameter.

\subsection{General sliceable rings}

The visibility of a general non-axisymmetric thick ring can only be computed after devising some criterion for deciding whether the image is indeed a ``ring''; otherwise, the intensity $I(x,y)$ may represent anything and will not necessarily have a visibility whose perturbative expansion in small $uw\ll1$ reduces to the universal form \eqref{eq:UniversalAmplitude} in the limit of zero width $w\to0$.
For this reason, in this work, we focus on thick rings that are \textit{sliceable}:
\begin{definition}[Sliceable ring]
A (possibly thick) ring is sliceable within the visibility baseline window $(u_-,u_+)$, if it can be decomposed (or ``sliced'') into a series of thin convex rings, each of which satisfies the universal regime conditions\footnote{Here, $w$ stands for the width of the thin ring slice.
In practice, it can almost always be treated as vanishingly small ($w\to0$).} $\max(1/d,1/\xi_\mathcal{I}) \ll u \ll 1/w$ and $\mathcal{I}_L\mathcal{I}_R>0$ for any $u\in(u_-,u_+)$.
\label{def:Slice}
\end{definition}
If a ring is sliceable, then the intensity near the center of the ring should be (at least approximately) zero, or else $1/d\ll u$ cannot hold for the innermost thin ring slice.
\cref{def:Slice} only gives the necessary condition for a ring to be sliceable.
In this work, we do not address the question of what conditions on an image are sufficient for it to be sliceable.
We do, however, present a numerical scheme in Sec.~\ref{sec:RadialQuantiles} that shows that GRMHD-simulated models of the time-averaged $n\ge1$ subrings of M87* at 230\,GHz are sliceable.\footnote{In general relativity, the $n\ge1$ subrings are confined to lensing bands \cite{gralla2020lensing}, so the intensity is zero inside (outside) the inner (outer) boundary of the lensing band, regardless of the plasma profile.}

\begin{figure}[t]
    \centering
    \includegraphics[width=\columnwidth]{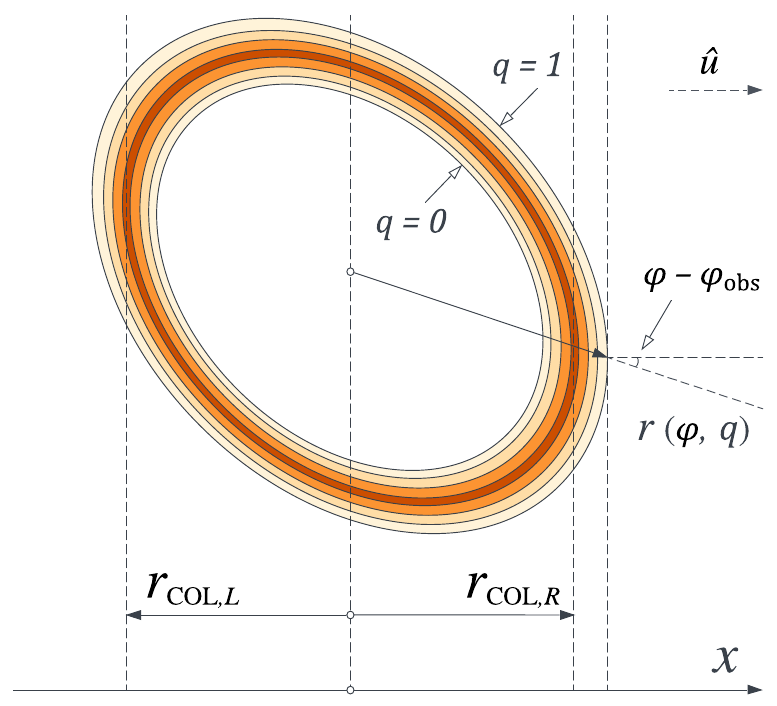}
    \caption{A sliceable thick ring can be decomposed into a series of convex thin rings $r(\varphi,q)$, labeled by an auxiliary variable $q\in[0,1]$.
    The visibility of each thin ring follows the universal  form \eqref{eq:UniversalAmplitude}, and the visibility of the thick ring can be calculated by summing up all of the thin ring contributions.
    The linear component in the oscillating phase of the visibility amplitude measures the sum of $r_L$ and $r_R$, which are the left and right \textit{center-of-light} radii of the ring, as defined in Eq.~\eqref{eq:rCOL}.}
    \label{fig:Slicing}
\end{figure}

Technically, a slicing scheme is defined via the family of closed convex curves that decomposes the thick ring into thin rings, as illustrated in Fig.~\ref{fig:Slicing}.
We take the origin to lie inside the ring and adopt the polar parameterization $r(\varphi,q)$, where $q\in[0,1]$ labels a family of nested curves with $r(\varphi,q_1)>r(\varphi,q_2)$ if $q_1>q_2$.
When projecting onto the axis $\bm{\hat{u}}\equiv (\cos {\varphi_{\rm obs}},\sin{\varphi_{\rm obs}})$, the left (right) edge need not be at $\varphi=\varphi_{\rm obs}+\pi$ ($\varphi=\varphi_{\rm obs}$), but rather corresponds to the angle $\varphi=\varphi_{L/R}$ that minimizes (maximizes) the projection $\br(\varphi,q)\cdot\bm{\hat{u}}$, which is also the angle where the normal vector to $\br(\varphi,q)$ is (anti-)parallel to $\bm{\hat{u}}$.
If a thick ring has \textit{surface} intensity $I(r,\varphi)$, then the thin ring slice between $q$ and $q+{\rm d}q$ has an effective \textit{line} intensity
\begin{align}
    \mathcal{I}(q,\varphi_{\rm obs})\equiv I_q(q,\varphi_{\rm obs})\,{\rm d}q,\label{eq:mathcalIq1}
\end{align}
At the edge points $\varphi=\varphi_{L/R}$, this becomes
\begin{align}
    \mathcal{I}(q,\varphi_{\rm obs})=I(r(\varphi,q),\varphi)\cos(\varphi-\varphi_{\rm obs})\frac{\partial r(\varphi,q)}{\partial q}{\rm d}q,\label{eq:mathcalIq2}
\end{align}
where the edge point angles $\varphi_{L/R}$ need to be solved for in terms of the given $r(\varphi,q)$ and $\varphi_{\rm obs}$ as described above, and $\varphi-\varphi_{\rm obs}$ is the angle between $\bm{\hat{r}}$ and the vector normal to the ring.
The curvature radius of a ring slice with polar parameterization $r_q(\varphi)=r(\varphi,q)$ is given by
\begin{align}
    \mathcal{R}(q,\varphi)=\frac{\left[r_q(\varphi)^2+r_q'(\varphi)^2\right]^{3/2}}{\left|r_q(\varphi)^2+2r_q'(\varphi)^2-r_q(\varphi)r_q''(\varphi)\right|}.
\end{align}

\subsection{Visibility of a sliceable ring}
\label{sec:VisibilitySlicing}

If a thick ring is sliceable, then each of its thin ring slices has a visibility of the form \eqref{eq:UniversalVisibility} in the universal regime \eqref{eq:UniversalRegime}.
Summing over the contributions from all the ring slices then produces the visibility of the thick ring.
Here, we assume that a viable slicing scheme exists and compute the visibility of the thick ring---we will explore possible strategies for slicing a ring in practice later.

In Sec.~\ref{sec:ThinConvexRing}, we saw that the visibility of a convex thin ring is dominated by the $|x-x_0|^{-1/2}$ singularities in its projected intensity that arise from its left and right edges.

\noindent A similar conclusion holds for a sliceable (convex) thick ring, except that each peak in its projected intensity now consists of a sum over the $|x-x_0|^{-1/2}$ weighted by the radial intensity profile.
Thus, the visibility of a sliceable thick ring can be calculated by adding the visibility of these two peaks, with each contributing a radial integral over the visibility \eqref{eq:HeavisideFT} of a single $|x-x_0|^{-1/2}$ singularity.

Fixing a projection direction $\bm{\hat{u}}$, for each thin ring slice labeled by $q$, we have an effective line intensity $I_q(q)\,{\rm d}q$ [defined in Eqs.~\eqref{eq:mathcalIq1}--\eqref{eq:mathcalIq2}], and two edge points $\varphi=\varphi_{L/R}$ where the curvature radius of the ring is $\mathcal{R}(q)=\mathcal{R}(\varphi,q)$ and its projected position is $x(q)$.\footnote{To simplify notation, we omit the dependence on $\varphi_{\rm obs}$, as well as on the subscripts $L$ or $R$ if an expression applies to both edges.}
We can then define the (one-side) \textit{center-of-light} radius of the thick ring as
\begin{align}
    \label{eq:rCOL}
    r_{\rm COL}=\alpha\frac{\int x(q)I_q(q)\sqrt{\mathcal{R}(q)}\,{\rm d}q}{\int I_q(q)\sqrt{\mathcal{R}(q)}\,{\rm d}q}
    >0,
\end{align}
where $\alpha=+1$ for the right side and $\alpha=-1$ for the left side.
The center-of-light diameter of the thick ring is then the sum of the left and right center-of-light radii,
\begin{align}
    \label{eq:dCOL}
    d_{\rm COL}\equiv r_{{\rm COL},L}+r_{{\rm COL},R}.\\
    \nonumber
\end{align}
With the local coordinate $r'$ defined by $x=\alpha(r_{\rm COL}+r')$ on each side, the visibility of the left or right peak in the projected intensity profile is (we use Eq.~(36) of Ref.~\cite{gralla2020measuring})
\begin{align}
    V_{L/R}(u)&\approx\frac{1}{\sqrt{u}}\int I_q(q)\sqrt{\mathcal{R}(q)}e^{i\alpha\frac{\pi}{4}}e^{-2\pi ixu}\,{\rm d}q\nonumber\\
    \label{eq:PeakVisibility}
    &\approx e^{i\alpha\frac{\pi}{4}}\frac{A_{L/R}(u)e^{-2\pi i\alpha r_{\rm COL}u}}{\sqrt{u}}e^{-i\alpha\phi_{L/R}},
\end{align}
where in the last step, we introduced new functions
\begin{align}
    \label{eq:EnvelopeAmplitude}
    A_{L/R}(u)&=\left|\int I_q(q)\sqrt{\mathcal{R}(q)}e^{-2\pi ir'u}\,{\rm d}q\right|,\\
    \label{eq:EnvelopePhase}
    \phi_{L/R}(u)&=\arctan{\frac{\int I_q(q)\sqrt{\mathcal{R}(q)}\sin(2\pi r' u)\,{\rm d}q}{\int I_q(q)\sqrt{\mathcal{R}(q)}\cos(2\pi r' u)\,{\rm d}q}}.
\end{align}
Using the moments $M_m$ of the ring profile,
\begin{align}
    \label{eq:Moments}
    M_m=\int I_q(q)\sqrt{\mathcal{R}(q)}\frac{(2\pi r')^m}{m!}\,{\rm d}q,
\end{align}
the first of which vanishes ($M_1=0$) by definition of $r_{\rm COL}$, we can expand the functions $A_{L/R}(u)$ and $\phi_{L/R}(u)$ as polynomials in $u$ with coefficients $a_m$ given in App.~\ref{app:Coefficients}:
\begin{widetext}
\begin{align}
    A_{L/R}(u)&=\sqrt{\left[\sum_{m=0}^\infty(-1)^mM_{2m}u^{2m}\right]^2+\left[\sum_{m=1}^\infty(-1)^mM_{2m+1}u^{2m+1}\right]^2}
    =\sum_{m=0}^\infty a_{2m}u^{2m},\nonumber\\
    \label{eq:EnvelopeExpansion}
    \phi_{L/R}(u)&=\arctan{\frac{\sum_{m=1}^\infty(-1)^mM_{2m+1}u^{2m+1}}{\sum_{m=0}^\infty(-1)^mM_{2m}u^{2m}}}
    =\sum_{m=1}^\infty a_{2m+1}u^{2m+1}.
\end{align}
\end{widetext}
The total visibility of a sliceable thick ring is the sum of the left and right peaks,
\begin{align}
    \label{eq:V2-thick-general}
    V(u)&=V_L(u)+V_R(u),\\
    |V(u)|^2&=\frac{A_L^2+A_R^2+2A_LA_R\sin(2\pi d_{\rm COL}u+\phi_L+\phi_R)}{u}.\nonumber
\end{align}
In the zero-width limit, all the $M_m$ and $a_m$ with $m>0$ vanish and we have $A_{L/R}=\mathcal{I}_{L/R}\sqrt{\mathcal{R}_{L/R}}$ and $\phi_{L/R}=0$, so that Eq.~\eqref{eq:V2-thick-general} reduces to Eq.~\eqref{eq:UniversalAmplitude}.
When the ring has finite width, we still have $M_1=a_1=0$ due to our choice of $r'$, such that the linear component of the oscillating phase only depends on the center-of-light diameter of the thick ring.
However, there are additional corrections to the oscillating envelope involving even powers of $u$, with the leading-order term proportional to $u^2$.
Similarly, the oscillating phase includes corrections in odd powers of $u$, with the leading-order term proportional to $u^3$.

When analyzing measured visibility data, these higher-order corrections can be separated from the linear term based on their different $u$-dependence.
If $I_q\sqrt{\mathcal{R}}\frac{\partial q}{\partial r'}$ is an even function of $r'$, then $\phi_{L/R}=0$ regardless of the ring width.\footnote{Once the convex thick ring is sliced into thin rings, there is at most one slice whose leftmost (rightmost) point is at $r'$, and so one can change the integration variable from $q$ to $r'$ in Eq.~\eqref{eq:Moments}.}
In other words, higher-order phase corrections only appear if the ring profile is \textit{radially nonsymmetric}.

In Eq.~\eqref{eq:EnvelopeExpansion}, we expanded $A_{L/R}$ and $\phi_{L/R}$ as power series in $u$.
In practice, we also truncate these infinite series after a few terms, which amounts to approximating the $\sin{2\pi r'u}$ and $\cos{2\pi r'u}$ terms in Eqs.~\eqref{eq:EnvelopeAmplitude}--\eqref{eq:EnvelopePhase} with polynomials.
Roughly speaking, for a thick ring of width $w$, this is a good approximation on baselines $u\lesssim1/w$, as opposed to the stricter condition $u\ll1/w$ required by the universal regime.
For $u\gg1/w$, the power series in Eq.~\eqref{eq:EnvelopeExpansion} diverge, although one can still evaluate $A_{L/R}$ and $\phi_{L/R}$ using the original integrals in Eqs.~\eqref{eq:EnvelopeAmplitude}--\eqref{eq:EnvelopePhase}.

\subsection{Circular slicing of a finite-width ring}

If the thick ring can be sliced into \textit{circular} thin rings, all within the universal regime, then the results from the previous subsection can be recovered from the circular thin ring calculations in Sec.~\ref{sec:ThinFourierRing}.
Indeed, let $C_{\rm e/o}(\rho)$ denote the ``radial density'' of $\mathcal{C}_{\rm e/o}$, so $\mathcal{C}_{\rm e/o}=C_{\rm e/o}(\rho)\,{\rm d}\rho$ for the annulus between $\rho$ and $\rho+{\rm d}\rho$.
By Eqs.~\eqref{eq:FourierVisibility}--\eqref{eq:EvenOdd}, we can then calculate the visibility of the thick ring by summing up the visibility of circular thin rings,
\begin{align}
    V(u)&=\frac{1}{\sqrt{u}}\bigg[\int C_{\rm e}(\rho)\cos\left(2\pi\rho u-\frac{\pi}{4}\right)\,{\rm d}\rho\nonumber\\
    &\qquad\quad+i\int C_{\rm o}(\rho)\sin\left(2\pi\rho u-\frac{\pi}{4}\right)\,{\rm d}\rho\bigg]\nonumber\\
    &=\frac{1}{\sqrt{u}}\bigg[\int I_L(\rho)\sqrt{\rho}\,e^{i\left(2\pi\rho u-\frac{\pi}{4}\right)}\,{\rm d}\rho\nonumber\\
    &\qquad\quad+\int I_R(\rho)\sqrt{\rho}\,e^{-i\left(2\pi\rho u-\frac{\pi}{4}\right)}\,{\rm d}\rho\bigg],
\end{align}
which agrees with Eq.~\eqref{eq:PeakVisibility}.   

While the circular slicing works for axisymmetric rings with $I(\rho,\varphi)=I_{\rho}(\rho)$, as well as for \textit{separable} rings with $I(\rho,\varphi)=I_{\rho}(\rho)I_{\varphi}(\varphi)$ if $I_{\varphi}(\varphi)$ is sufficiently smooth, it is not always a viable strategy for generic rings since the circular thin ring slices may not satisfy the universal regime conditions in Eq.~\eqref{eq:UniversalRegime}.
As an example, consider a thick ring with nonzero intensity within the elliptical band $\gamma_1^2\leq x^2/a^2+y^2/b^2\leq\gamma_2^2$ ($a>b>0$, $\gamma_2>\gamma_1>0$) and zero otherwise.
Then for $\gamma_1 b<r<\gamma_1 a$, the circle of radius $r$ will not lie entirely within the thick ring.
It will thus have $\mathcal{I}_L=\mathcal{I}_R=0$ as well as discontinuities in $\mathcal{I}(s)$, and hence its visibility cannot be evaluated using the universal regime solution.
However, such a thick ring is actually \textit{sliceable}, as one can slice it into elliptical thin rings $x^2/a^2+y^2/b^2=\gamma^2$ with $\gamma_1\leq\gamma\leq\gamma_2$, for which the universal regime conditions hold.

\subsection{Non-circular slicing using radial intensity quantiles}
\label{sec:RadialQuantiles}

The elliptical band example suggests that, in order to ensure the universal regime criteria are met by each thin ring slice, the slicing should follow the intensity profile of the thick ring.
This observation motivates the following slicing scheme based on radial intensity quantiles.
Along each ray of fixed $\varphi$, we compute the $q^\text{th}$ quantiles $r_q(\varphi)$ of the radial intensity $I(r,\varphi)$ by solving the equation
\begin{align}
    \frac{\int_0^{r_q(\varphi)}I(r',\varphi)\,{\rm d}r'}{\int_0^\infty I(r',\varphi)\,{\rm d}r'}=q.
\end{align}
The $q^\text{th}$ thin ring slice then consists of the $q^\text{th}$ quantiles at every $\varphi$, or in other words:
\begin{align}
    r(\varphi,q)\equiv r_q(\varphi).
\end{align}
Such a construction guarantees that each thin ring slice has a smooth angular profile, as long as the intensity profile $I(r,\varphi)$ of the thick ring is sufficiently smooth.

With this quantile-based scheme, the exact slicing of the thick ring does depend on the choice of origin for the coordinate system, which we typically take to be the origin of the Bardeen coordinates \cite{bardeen1973timelike}.
We stress that the slicing scheme presented here is specifically designed for \textit{simulated images}, for which the Bardeen coordinates of each pixel are always known.

By contrast, in real observations, the ``true'' black hole position is not known, but this should not be a concern since one works directly in visibility space.
Nevertheless, our experiments indicate that with a slight shift of the origin---e.g., to Bardeen coordinates $(x,y)=(0.5M,0)$---the visibility amplitude predicted by our perturbative calculation remains unchanged.

\section{Numerical experiments}
\label{sec:Numerics}

\begin{figure}[!ht]
    \centering
    \includegraphics[width=\columnwidth]{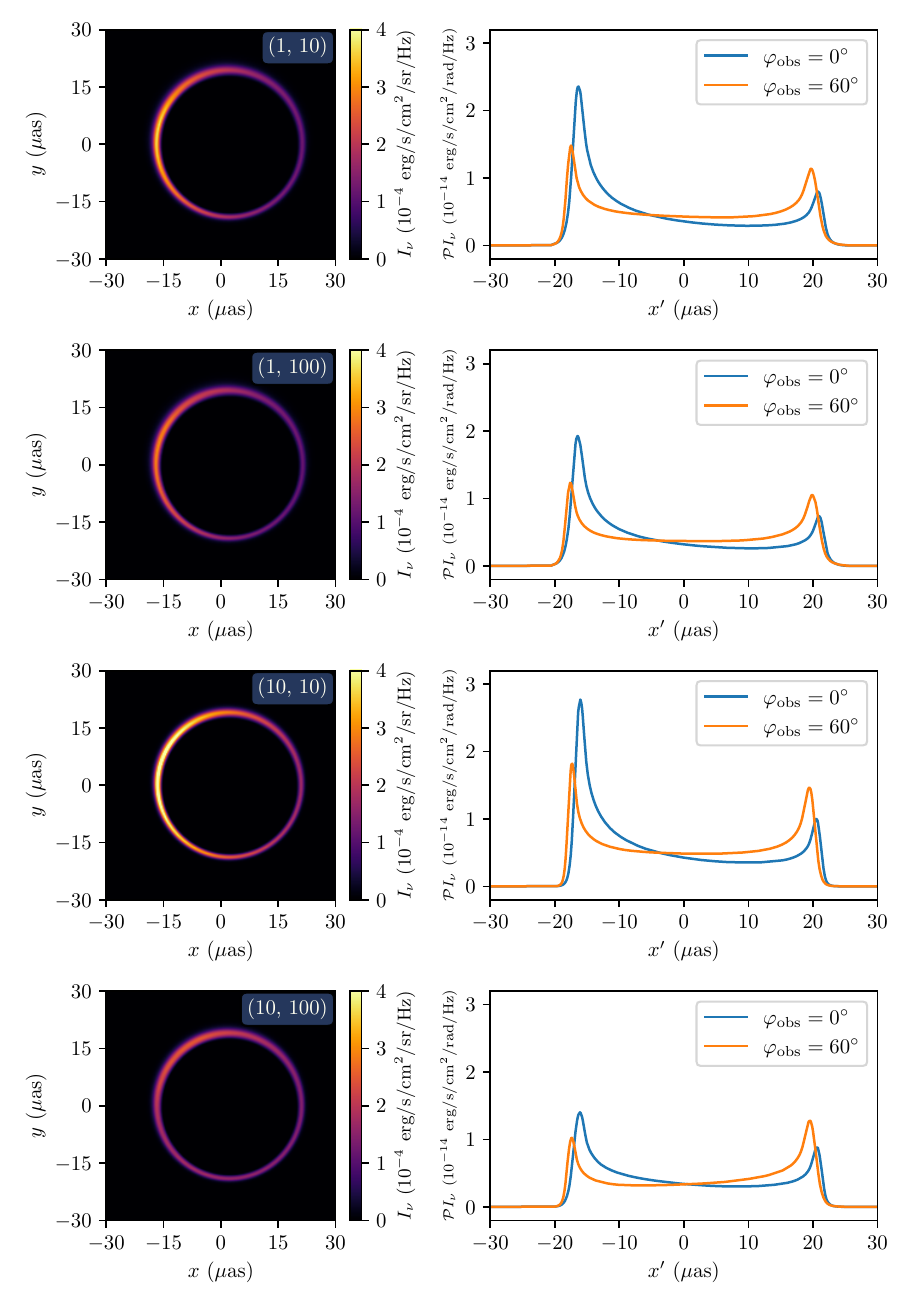}
    \caption{Time-averaged images of the first photon ring of M87* from the MAD GRMHD simulation, and its projections along $\varphi_{\rm obs}=0^{\circ}$ and $\varphi_{\rm obs}=60^{\circ}$; different electron temperature models are shown, with $(R_{\rm low},R_{\rm high})$ set to $(1,10)$, $(1,100)$, $(10,10)$ and $(10,100)$.
    As shown in Figs.~\ref{fig:1vs5} and \ref{fig:phase-mad-0}, these four rings are all \textit{sliceable} and their visibility amplitudes agree with the analytic prediction in Eq.~\eqref{eq:V2-thick-general}.}
    \label{fig:img-mad}
\end{figure}

\begin{figure*}[!ht]
    \centering
    \includegraphics[width=\textwidth]{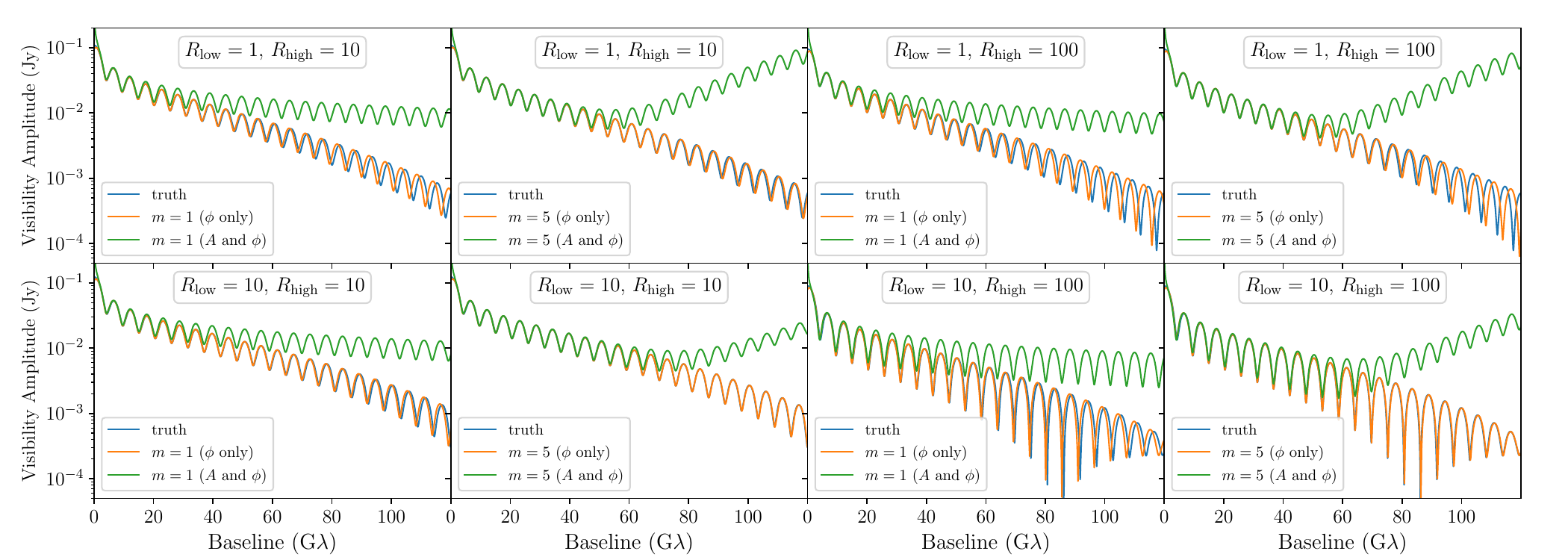}
    \caption{The visibility amplitude along $\varphi_{\rm obs}=0^{\circ}$ for the MAD simulation with four different electron temperature models.
    We compare the visibility calculated with an FFT of the projected intensity (truth), perturbation theory up to first order ($m=1$, the universal regime behavior), and perturbation theory up to fifth order ($m=5$).
    We also compare the case where both the envelope and the phase are calculated using perturbation theory ($A$ and $\phi$), with the case where the phase is calculated perturbatively and the envelope is directly fitted from the true $|V(u)|$ ($\phi$ only).
    We find that fifth-order perturbation theory predicts the visibility more accurately than the universal regime formula, while the phase is generally perturbative up to longer baselines than the envelope.
    A direct comparison of the phase error $\Delta\phi\equiv\phi_m-\phi_{\rm true}$ is shown in Fig.~\ref{fig:phase-mad-0}.}
    \label{fig:1vs5}
\end{figure*}

\begin{figure*}[!ht]
    \centering
    \includegraphics[width=\textwidth]{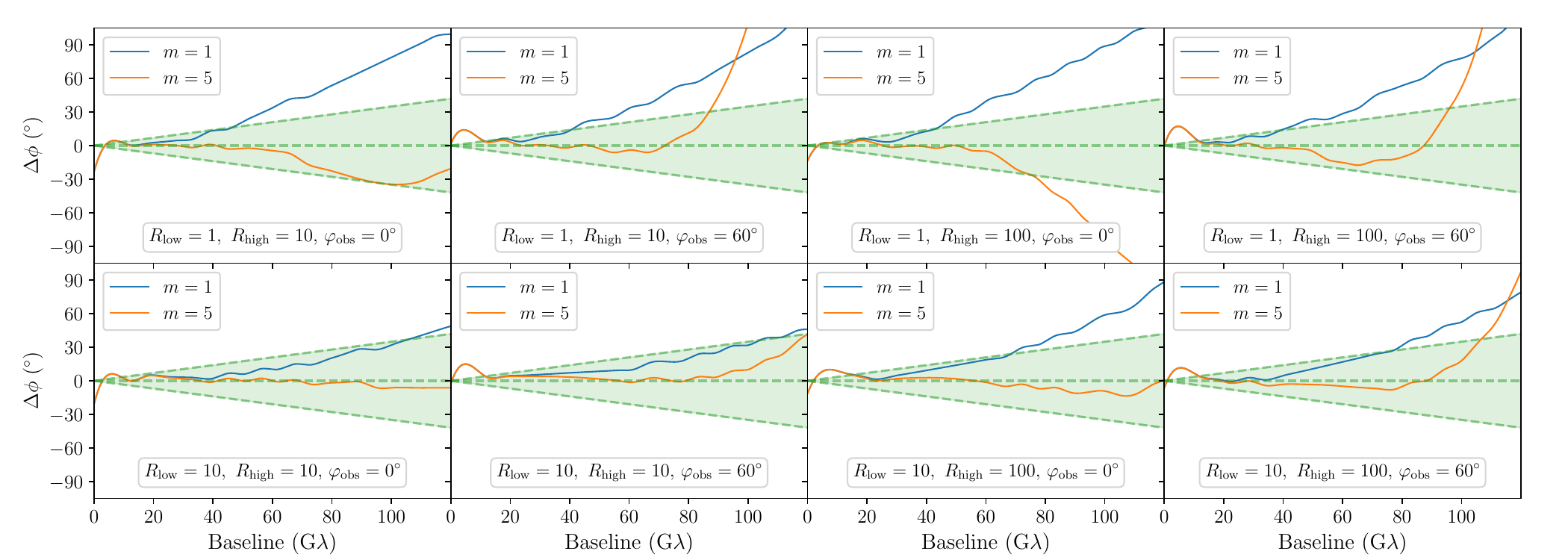}
    \caption{The difference in phase $\Delta\phi\equiv\phi_m-\phi_{\rm true}$ between the perturbative calculation carried out to $m^\text{th}$ order and the phase extracted from the true visibility amplitude using the method in App.~\ref{app:EnvelopePhase}, for the MAD simulation.
    The green bands indicate the phase shift due to a $0.2\,\mu$as difference in the center-of-light diameter of the ring.
    The $m=5$ prediction has a more accurate phase than the linear approximation, with an expected diameter measurement error of less than $0.2\,\mu$as at $u\lesssim80\,{\rm G}\lambda$, for all the four electron temperature models studied in this work.}
    \label{fig:phase-mad-0}
\end{figure*}

\begin{figure*}[!ht]
    \centering
    \includegraphics[width=\textwidth]{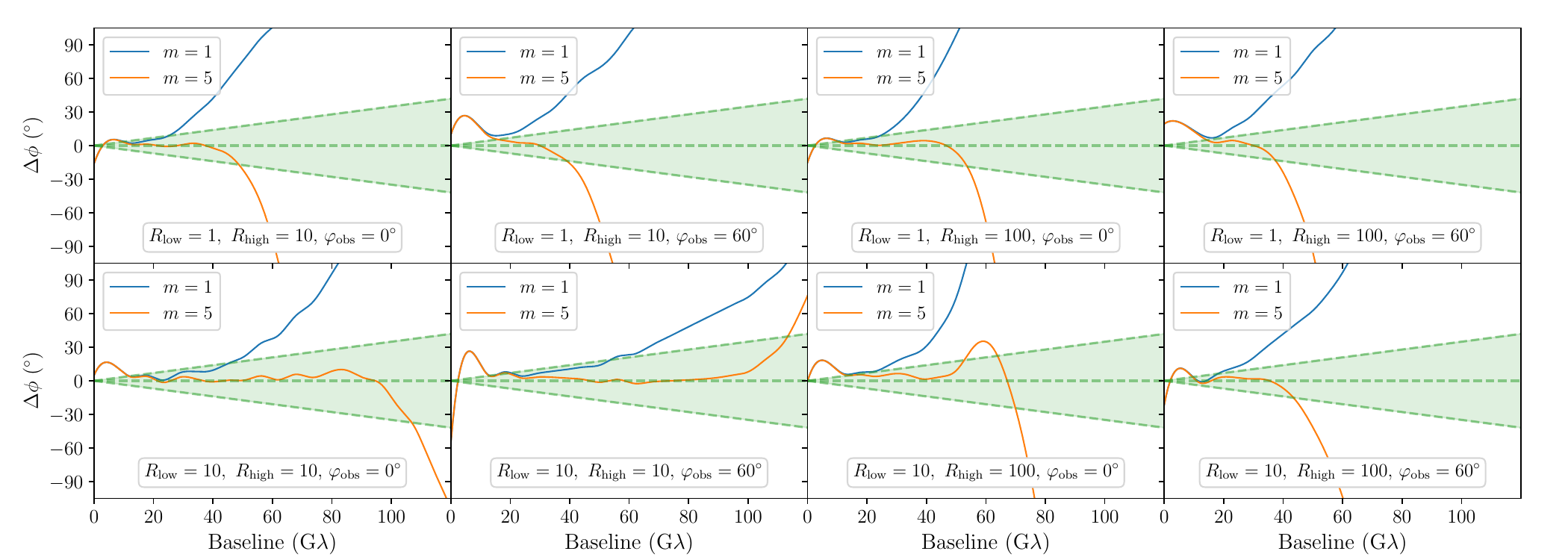}
    \caption{Similar to Fig.~\ref{fig:phase-mad-0}, but for the SANE simulation.
    The $m=5$ formula still outperforms $m=1$ in terms of the phase prediction, although we find that the first photon ring is generally thicker and therefore becomes non-perturbative at shorter baselines for the SANE simulation compared with the MAD one.}
    \label{fig:phase-sane-0}
\end{figure*}

\begin{figure*}[!ht]
    \includegraphics[width=\textwidth]{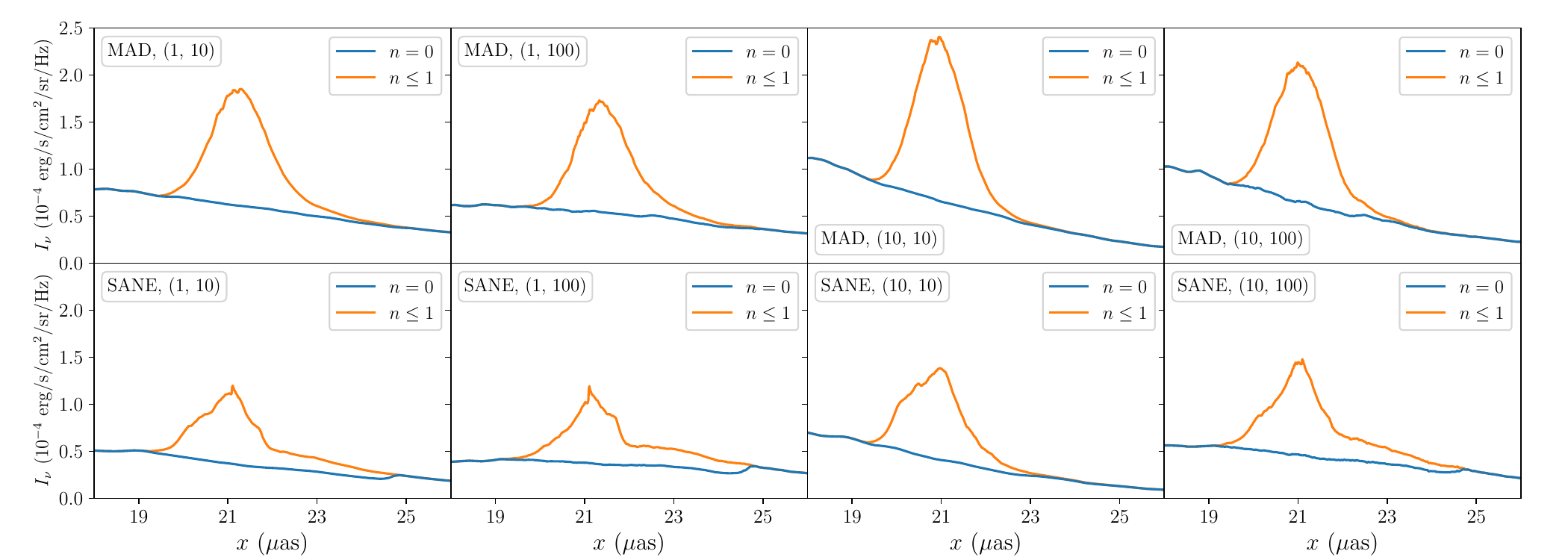}
    \caption{Slices of the photon ring images along $y=0$ for the different GRMHD and electron temperature models.
    In each panel, we show the $n=0$ subimage (blue) and the total $n\leq1$ subimage (orange), while their difference stands for the $n=1$ subimage.
    The SANE models with $(R_{\rm low},R_{\rm high})$ set to $(1,10)$, $(1,100)$ and $(10,100)$ have a noticeably heavier tail than the other models, which explains why the perturbative expansion performs less well for the first photon ring in these models.}
    \label{fig:slice-mad-sane}
\end{figure*}

We test our scheme against two numerical simulations of supermassive black hole accretion to gauge how well it performs in the face of complicated accretion physics.

We use the {\tt athenak} code \cite{stone2024athenak} for the fluid simulations.
This code solves the hyperbolic, conservative equations of ideal GRMHD on a logically Cartesian, statically refined mesh in Cartesian Kerr-Schild coordinates.
The domains in the simulations extend to $\pm1024\,GM/c^2$, have eight total refinement layers, and feature a resolution of $16$ zones per $GM/c^2$ within a $\pm8\,GM/c^2$ cube centered on the black hole.

We perform two simulations in order to cover both the magnetically arrested disk (MAD) as well as the standard and normal evolution (SANE) accretion states.
Both simulations assume a black hole spin of $a/M=93.75\%$ and are initialized from a Fishbone-Moncrief \cite{fishbone1976relativistic} torus configuration, with the inner edge at $r=20\,GM/c^2$ and the pressure maximum at $r=41\,GM/c^2$.
We set the fluid adiabatic index to $13/9$ globally across the flow.

We raytrace the simulations with a modified version of the \texttt{blacklight} code \cite{white2022blacklight} that allocates more pixels within the lensing bands where small scale demagnified structures are expected.
The details of our nested ray tracing scheme can be found in \cite{jia2024modeling}.
For post-processing, we make use of the $(R_{\rm low},R_{\rm high})$ model \citep{moscibrodzka2016general} to compute the electron temperature $T_e$ from GRMHD fluid pressure $p$, density $\rho$ and plasma $\beta$,
\begin{align}
    T_e&=\frac{2T_{\rm fluid}}{1+R},\nonumber\\
    \text{where}\quad T_{\rm fluid}&\equiv\frac{m_pp}{2k_B\rho},\nonumber \\
    R &\equiv\frac{T_p}{T_e}=\frac{\beta^2}{1+\beta^2}R_{\rm high}+\frac{1}{1+\beta^2}R_{\rm low},\nonumber \\
    \label{eq:R}
    \beta&\equiv \frac{8\pi p}{B^2}.
\end{align}
The camera lies at $r_0=100\,r_g$, $\theta_0=163^{\circ}$ and $\phi_0=0^{\circ}$.
We set the black hole mass to $M=6.5\times10^9\,M_{\odot}$ and use $1M=3.8\,\mu$as to convert from Bardeen's coordinates to angular distances, corresponding to the mass-to-distance ratio for M87*.
We get the image of the $n^\text{th}$ photon ring by taking the difference between the ray tracing starting from the $(n+1)^\text{th}$-to-last turning point\footnote{Here, the turning point is defined as the point with ${\rm d}z/{\rm d}\lambda=0$, where $z$ is the Cartesian Kerr-Schild coordinate and $\lambda$ is the affine parameter of the geodesic.
We start from the horizon or infinity if there are fewer than $n+1$ turning points on the geodesic.}
on the geodesic, and the ray tracing starting from the $n^\text{th}$-to-last turning point.
Then, we average over 500 snapshots at a cadence of $10M$ and calculate the ``true'' visibility using the Fast Fourier Transform (FFT).\\

The averaged images and projected intensity profiles corresponding to $(R_{\rm low},R_{\rm high})$ equal to $(1,10)$, $(1,100)$, $(10,10)$ and $(10,100)$ are shown in Figs.~\ref{fig:img-mad} and \ref{fig:img-sane} for the MAD and SANE simulations, respectively.
Though the exact electron temperature prescription for M87* is still uncertain, we note that the two intermediate models with $(1,100)$ and $(10,10)$ are in generally better agreement with the current EHT observations (see, e.g., Ref.~\cite{eht2019m87v}).

Rather than treating the $a_m$ as free parameters and fitting the visibility amplitude with Eq.~\eqref{eq:V2-thick-general}, we directly apply our slicing scheme in Sec.~\ref{sec:RadialQuantiles} to the coherently averaged images, and evaluate the $a_m$ from the moments $M_m$ of the ring profile.
As further discussed in App.~\ref{app:range}, instead of calculating the moments \eqref{eq:Moments} with the integral running over the whole range $[0,1]$ for $q$, we truncate it to $[0.1\%,99.9\%]$ so as to mitigate the instabilities caused by the tail of the integrand.
We then compare the $|V(u)|$ calculated via Eq.~\eqref{eq:V2-thick-general} up to the $m^\text{th}$ order and the actual visibility amplitude of the image in Fig.~\ref{fig:1vs5} for the MAD simulation, with $m=1$ representing the universal regime formula.
Since in practice, we use the perturbative phase $\phi$ for the analysis of visibility data, as discussed in Sec.~\ref{sec:Discussion}, we include a case where both the envelope and phase are calculated with perturbation theory, and also a case with only the phase computed perturbatively.
For the latter, we use the method in Sec.~\ref{app:EnvelopePhase} to numerically decompose the true visibility into $A(u)$, $B(u)$ and $\phi(u)$ as follows:
\begin{align}
    \label{eq:V2ABsinphi}
    |V(u)|^2=A(u)+B(u)\sin{\phi(u)}.
\end{align}
The envelope of $|V(u)|^2$ can then be obtained from $A(u)$ and $B(u)$.
As shown in Fig.~\ref{fig:1vs5}, the peaks and valleys of the perturbative $m=1$ approximation to $|V(u)|$ misalign with the true visibility due to the error in the phase at $u\sim100\,{\rm G}\lambda$, which is improved by the $m=5$ model.
Moreover, the phase is more amenable than the envelope to perturbative treatment, since for $m=5$, the envelope deviates from the truth at around 40-50$\,{\rm G}\lambda$, even though the phase is still accurate up to about 80-100$\,{\rm G}\lambda$.

We also compare the true $\phi(u)$ to the perturbatively predicted phase at $m^\text{th}$ order with $m=1$ and $m=5$, and plot their difference in Fig.~\ref{fig:phase-mad-0} for the MAD simulation, which clearly shows that $m=5$ improves the accuracy of the $\phi$ modeling.
An error in the center-of-light diameter leads to a phase error $\Delta\phi\equiv\phi_m-\phi_{\rm true}$ that is linearly dependent on $u$ [see Eq.~\eqref{eq:V2-thick-general}], and the boundary of the green bands in Fig.~\ref{fig:phase-mad-0} represents the $\Delta\phi$ corresponding to a $0.2\,\mu$as difference in the diameter.
Therefore, should there be no other forms of noise, the systematic error of the center-of-light diameter measurement will be less than $0.2\,\mu$as if (and only if) $\Delta\phi$ falls within the green bands.
We find that the phase from the $m=5$ model closely matches the truth at shorter baselines $u\lesssim50\,{\rm G}\lambda$, and the expected diameter measurement error remains less than 0.2$\,\mu$as at $u\lesssim80\,{\rm G}\lambda$ for all four of our electron temperature models.
By contrast, with the $m=1$ model, we expect a systematic error of 0.2$\,\mu$as in the center-of-light diameter, even when the measurement is conducted on shorter baselines $u\sim40\,{\rm G}\lambda$.

In Fig.~\ref{fig:phase-sane-0}, we show the analogous results for the SANE simulation, in which we find that the ring is generally thicker (see Fig.~\ref{fig:slice-mad-sane}), therefore becoming non-perturbative on shorter baselines.
More detailed discussion regarding the SANE results can be found in App.~\ref{app:SANE}.

\section{Discussion}
\label{sec:Discussion}

In this paper, we investigate the visibility of black hole photon rings beyond the universal regime by accounting perturbatively for the effects of finite ring width.
Here, we focus on \textit{sliceable} thick rings (\cref{def:Slice}), which can be decomposed into multiple thin ring slices such that the visibility of the thick ring is the sum of the visibilities of the thin rings.
In this paper, we do not determine the sufficient conditions on a ring image for it to be sliceable.
However, we do show that photon ring images ray traced from GRMHD simulations can be sliced using the scheme in Sec.~\ref{sec:RadialQuantiles} based on radial intensity quantiles.

The main contributions of this work are that:
\begin{enumerate}
    \item We show that the exact diameter measured for \textit{sliceable} thick rings from the oscillation of their visibility amplitude is precisely the center-of-light diameter defined in Eq.~\eqref{eq:dCOL}.
    \item We compute higher-order corrections (polynomial in $u$) to the visibility beyond the universal regime.
    We obtain baseline-dependent corrections $\propto u^3$, $u^5$, etc., to the oscillation in the visibility amplitude.
    Incorporating these higher-order terms is crucial for accurately inferring the part of the oscillation phase that is linear in the baseline $u$.
    It is this linear term that encodes the diameter of the ring, which in turn depends on the black hole mass and spin.
\end{enumerate}

The panels in Figs.~\ref{fig:phase-mad-0} and \ref{fig:phase-sane-0} give an intuitive estimate of the systematic error introduced by the thickness of the $n=1$ photon ring in the inference of its center-of-light diameter.
We find that this error stays below 0.2$\,\mu$as for $u\lesssim80\,{\rm G}\lambda$ in MAD simulations and for $u\lesssim40\,{\rm G}\lambda$ in SANE simulations, provided one includes the additional perturbative corrections to the oscillating phase of the visibility amplitude calculated here.
As a comparison, the typical width $w$ of the first lensing band for M87* is about $6\,\mu$as given our assumed mass-to-distance ratio, corresponding to a baseline length of $u\sim 1/w=34.4\,{\rm G}\lambda$.

\noindent This roughly sets a lower bound (among different plasma prescriptions) on the baseline $u$ past which we expect our perturbative approach to breakdown.
We find that the MAD models remain perturbative up to significantly longer baselines because the $n=1$ image is signficantly narrower than the total width of the first lensing band, which is not as true for the SANE models.

Recent work \cite{cardenas2023prediction} exploring diameter-fitting for thick rings from their interferometric visibility, and applied the fitting procedure previously derived for a zero-width ring \cite{paugnat2022photon} without theoretically justifying the unbiasedness of the method when applied to thick rings.
This approach results in different diameters when fitting within different baseline windows, likely because it does not account for the higher-order corrections to the phase of the visibility amplitude discussed herein.

The visibility of a ring fails to display the universal regime behavior \eqref{eq:UniversalAmplitude} if any of the conditions in Eq.~\eqref{eq:UniversalRegime} are violated.
Our analysis only addresses the change in the visibility of the ring due to finite ring width when the condition $u\ll 1/w$ is not met.
Our method may still fail (1) if the thick ring is not sliceable, that is, if it cannot be decomposed into multiple thin rings for which Eq.~\eqref{eq:UniversalRegime} holds, or (2) if the ring is sliceable but the slicing is not perturbative, meaning that the polynomial expansion in Eq.~\eqref{eq:EnvelopeExpansion} fails at large $u$ of interest.
It is not clear how to extend our perturbative calculation of the visibility when $u \gg 1/d$ or $u \gg 1/\xi_I$ does not hold, since the visibility then depends on the global intensity profile of the ring (rather than being dominated by its left and right edges).

We find that the inclusion of polynomial corrections up to fifth order to the oscillating visibility amplitude is sufficient for measurements on baseline lengths $u\lesssim 40 {\rm G}\lambda$ likely achievable even with space-VLBI in the near term.
However, our perturbative approach inevitably breaks down on yet-longer baselines $u\gtrsim 1/w$, as seen in Figs.~\ref{fig:phase-mad-0}, \ref{fig:phase-sane-0} and \ref{fig:phase-mad-2}.

It may be possible to further extend the convergence interval by, e.g., employing a different slicing scheme or rearranging higher-order terms via Pad\'e resummation (see, e.g., Ref.~\cite{seljak2015halo}).
Empirically, we find that if the ring is either nonsliceable or nonperturbative, then it is typically not possible to fit the oscillating phase of its visibility amplitude with a simple polynomial in $u$.
This observation suggests that, in practice, if one can measure the visibility within a baseline window that is sufficiently large to separate the different $u$-dependent terms, and one finds that the model achieves a good fit to the data, then it is likely that the ring diameter inferred from such a fit is correct and unbiased.
This conjecture will be the subject of more detailed tests on simulated data.

There are additional effects that should be accounted for before the models proposed herein can be applied to future observations; several of these will be addressed in a companion paper \cite{jia2024modeling}.

First, while we calculated the visibility amplitude of a single sliceable thick ring in Sec.~\ref{sec:ThickRings}, realistic black hole images include multiple nested photon rings due to the lensing of Kerr spacetime \cite{gralla2020lensing}, which in principle should be modeled as the sum of multiple sliceable rings.
But, since these rings dominate the visibility on different baselines, the total visibility on a given range of baselines can typically be well-approximated with that of a single ring \cite{johnson2020universal}, provided one excludes the transition regions between regimes dominated by different rings \cite{paugnat2022photon,cardenas2024assessing}.

Second, due to the difficulties inherent in VLBI phase calibration, we can only measure the averaged visibility amplitude of the image (the \textit{incoherent} average), and not the visibility amplitude of the time-averaged image (the \textit{coherent} average) studied in this work.
Even though an incoherent average changes the envelope of the visibility amplitude (relative to a coherent average), nevertheless the shift of the phase becomes negligible after averaging over sufficiently long timescales, such that one can still infer the ring diameter from the phase.

In addition, to obtain the visibility across different baselines, one needs to vary the orbit of a space-orbiting satellite or sample the visibility at different frequencies.
In the latter case, one would need to model the frequency dependence of the image.
Encouragingly, however, our analysis suggests that within the frequency window likely to be explored by forthcoming space-VLBI missions (e.g., between 230\,GHz and 345\,GHz), the black hole image only exhibits minor variation, while the center-of-light diameter $d_{\rm COL}$ and the coefficients $a_m$ can be treated as either constant or linearly dependent on the observing frequency, thereby enabling a straightforward modeling of the frequency dependence of the image.

\section*{Acknowledgements}

We thank Alejandro C\'ardenas-Avenda\~no, Andrew Chael, and Sihao Cheng for helpful discussions.
The work presented in this article was performed on computational resources managed and supported by Princeton Research Computing, a consortium of groups including the Princeton Institute for Computational Science and Engineering (PICSciE) and the Office of Information Technology's High Performance Computing Center and Visualization Laboratory at Princeton University.
This work was supported in part by a Simons Investigator Grant from the Simons Foundation (EQ) and by NSF grants AST-2307888 and PHY-2340457 (AL).
It greatly benefited from AL and EQ's stay at the Aspen Center for Physics, which is supported by National Science Foundation grant PHY-15 2210452, and by EQ's stay at the Kavli Institute for Theoretical Physics, supported by NSF PHY-2309135.

\bibliography{apssamp}

\clearpage

\onecolumngrid

\appendix

\section{Visibility of a circular thin ring}
\label{app:CircularThinRing}

In this section, we demonstrate how to derive Eq.~\eqref{eq:FourierVisibility} from Eq.~\eqref{eq:BesselExpansion}.
When $\pi du\gg m$, the $J_m(\pi du)$ terms in Eq.~\eqref{eq:BesselExpansion} can be approximated with trigonometric functions using the large-argument approximation of Bessel functions, namely $J_m(z)\sim\sqrt{\frac{2}{\pi z}}\cos\left(z-\frac{m\pi}{2}-\frac{\pi}{4}\right)$ for $z\gg m$.
Note that $(-i)^{-m}=(-1)^m\times(-i)^{m}$, $J_{-m}(\pi du)=(-1)^m\times J_m(\pi du)$, and for real ring images, $\text{Re}\,\beta_{-m}=\text{Re}\,\beta_{m}$, $\text{Im}\,\beta_{-m} =(-1)\times\text{Im}\,\beta_{m}$. 
Therefore, we have\footnote{Our Eqs.~\eqref{eq:FourierVisibility} and \eqref{eq:EvenOdd} are equivalent to Eq.~(20) of \cite{johnson2020universal}, but are written in a form that is more convenient for our discussion here.}
\begin{align}
    V(u)&=\beta_0 J_0(\pi du)+2\sum_{k=1}^{\infty}(-1)^k\,\text{Re}\,\beta_{2k}\,J_{2k}(\pi du)+(-2i)\sum_{k=0}^{\infty}(-1)^k\,\text{Re}\,\beta_{2k+1}\,J_{2k+1}(\pi du)\nonumber\\
    &\approx\frac{1}{\pi}\sqrt{\frac{2}{du}} \biggl[\beta_0\cos\left(\pi du-\frac{\pi}{4}\right)+2\sum_{k=1}^{\infty}\text{Re}\,\beta_{2k}\,\cos\left(\pi du-\frac{\pi}{4}\right)\nonumber\\
    &\qquad\qquad\qquad\qquad\qquad\qquad\qquad+(-2i)\sum_{k=0}^{\infty}\text{Re}\,\beta_{2k+1}\,\sin\left(\pi du-\frac{\pi}{4}\right)\biggr]\nonumber\\
    \label{eq:vu-thin-fourier}
    &=\frac{1}{\sqrt{u}}\left[\mathcal{C}_{\rm e}\cos\left(\pi du-\frac{\pi}{4}\right)+i\,\mathcal{C}_{\rm o}\sin\left(\pi du-\frac{\pi}{4}\right)\right],\\
    \text{where}\quad\mathcal{C}_{\rm e}&=\frac{1}{\pi}\sqrt{\frac{2}{d}}\left[\beta_0+2\sum_{k=0}^{\infty}\text{Re}\,\beta_{2k}\right],\quad
    \mathcal{C}_{\rm o}=-\frac{2}{\pi}\sqrt{\frac{2}{d}}\sum_{k=0}^{\infty}\text{Re}\,\beta_{2k+1}.
\end{align}
The $\text{Im}\,\beta_{m}$ terms vanish in the summation, since the projection of a sine ring along the $x$-axis is zero.
Taking the amplitude of Eq.~\eqref{eq:vu-thin-fourier} then yields Eq.~\eqref{eq:FourierVisibility}.

\clearpage

\section{Coefficients \texorpdfstring{$a_m$}{a(m)}}
\label{app:Coefficients}

Here, we list the expressions for the expansion coefficient $a_m$ introduced Eq.~\eqref{eq:EnvelopeExpansion}, up to $m=13$:
\begin{align}
    a_0&=M_0,\quad
    a_1\equiv0,\quad
    a_2=-M_2,\quad
    a_3=-\frac{M_3}{M_0},\quad
    a_4=M_4,\quad
    a_5=-\frac{M_2M_3}{M_0^2}+\frac{M_5}{M_0},\quad
    a_6=\frac{M_3^2}{2M_0}-M_6,\nonumber\\
    a_7&=-\frac{M_2^2M_3}{M_0^3}+\frac{M_3M_4+M_2M_5}{M_0^2}-\frac{M_7}{M_0},\quad
    a_8=\frac{M_2M_3^2}{2M_0^2}-\frac{M_3M_5}{M_0}+M_8,\nonumber\\
    a_9&=-\frac{M_2^3M_3}{M_0^4}+\frac{M_3^3+3M_2^2M_5+6M_2M_3M_4}{3M_0^3}-\frac{M_4M_5+M_3M_6+M_2M_7}{M_0^2}+\frac{M_9}{M_0},\nonumber\\
    a_{10}&=\frac{M_2^2M_3^2}{2M_0^3}-\frac{M_3^2M_4+2M_2M_3M_5}{2M_0^2}+\frac{M_5^2+2M_3M_7}{2M_0}-M_{10},\nonumber\\
    a_{11}&=-\frac{M_2^4M_3}{M_0^5}+\frac{M_2M_3^3+M_2^3M_5+3M_2^2M_3M_4}{M_0^4}-\frac{M_2^2M_7+M_3^2M_5+M_3M_4^2+2M_2M_4M_5+2M_2M_3M_6}{M_0^3}\nonumber\\
    &\quad+\frac{M_5M_6+M_4M_7+M_3M_8+M_2M_9}{M_0^2}-\frac{M_{11}}{M_0},\nonumber\\
    a_{12}&=\frac{M_2^3M_3^2}{2M_0^4}-\frac{M_3^4+8M_2M_3^2M_4+8M_2^2M_3M_5}{8M_0^3}+\frac{M_2M_5^2+M_3^2M_6+2M_3M_4M_5+2M_2M_3M_7}{2M_0^2}\nonumber\\
    &\quad-\frac{M_5M_7+M_3M_9}{M_0}+M_{12},\nonumber\\
    a_{13}&=-\frac{M_2^5M_3}{M_0^6}+\frac{M_2^4M_5+2M_2^2M_3^3+4M_2^3M_3M_4}{M_0^5}\nonumber\\
    &\quad-\frac{M_2^3M_7+M_3^3M_4+3M_2M_3^2M_5+3M_2M_3M_4^2+3M_2^2M_4M_5+3M_2^2M_3M_6}{M_0^4}\nonumber\\
    &\quad+\frac{M_2^2M_9+M_4^2M_5+M_3^2M_7+M_3M_5^2+2M_3M_4M_6+2M_2M_5M_6+2M_2M_4M_7+2M_2M_3M_8}{M_0^3}\nonumber\\
    &\quad-\frac{M_6M_7+M_5M_8+M_4M_9+M_3M_{10}+M_2M_{11}}{M_0^2}+\frac{M_{13}}{M_0}.
\end{align}

\clearpage

\section{Comparison between different perturbative orders}

\begin{figure*}[!ht]
    \centering
    \includegraphics[width=\textwidth]{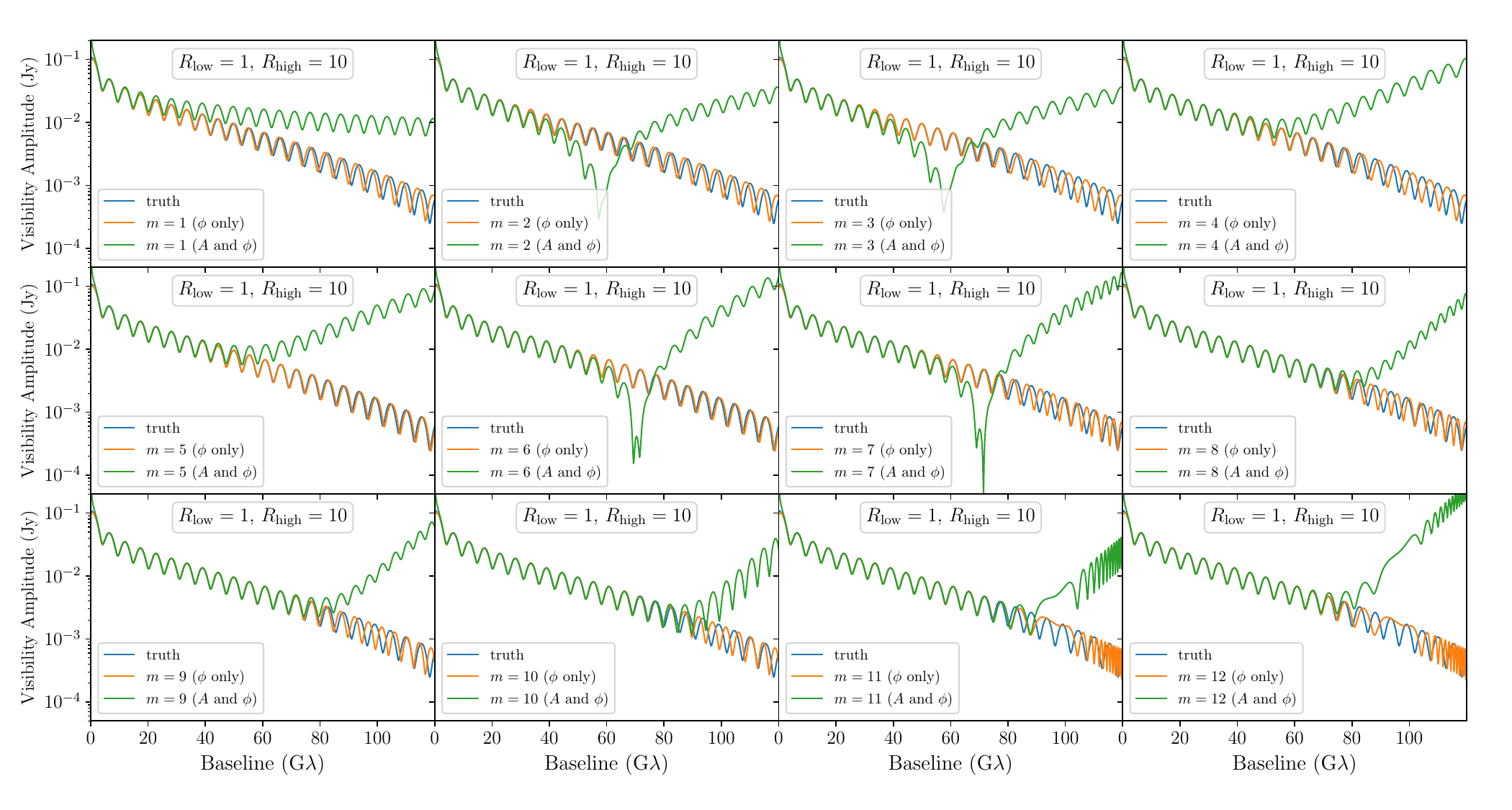}
    \caption{Comparison of the visibility amplitude for $(R_{\rm low},R_{\rm high})=(1,10)$ and $\varphi_{\rm obs}=0^{\circ}$ calculated at different perturbative orders.
    Generally, we find $m=5$ leads to the best results up to $100\,{G}\lambda$ for the first photon ring of M87*.
    The drop of the green curve at around $70\,{G}\lambda$ in some panels is due to the breakdown of perturbation theory, which predicts negative $A_{L/R}$.
    Note that $m=2i$ and $m=2i+1$ have the same perturbative $A_{L/R}$, while $m=2i$ and $m=2i-1$ share the same perturbative $\phi_{L/R}$.}
    \label{fig:m}
\end{figure*}

In Fig.~\ref{fig:m}, we compare the perturbative solution to the visibility amplitude up to different orders.
Although $m\ge8$ improves the envelope prediction up to about $80\,{\rm G}\lambda$, $m=5$ offers the most accurate results for the phase.
Since in practice, the envelope will be modeled with smooth spline functions instead of perturbative polynomials due to the noise from incoherent averaging \cite{jia2024modeling}, we recommend choosing $m=5$ for a better modeling of the phase.

\section{The envelope and phase of \texorpdfstring{$|V(u)|^2$}{the visibility amplitude}}
\label{app:EnvelopePhase}

We use the following approach to numerically decompose $|V(u)|^2$ into $A(u)$, $B(u)$ and $\phi(u)$ as in Eq.~\eqref{eq:V2ABsinphi}, which assumes that $A(u)$, $B(u)$ and $\phi(u)$ are all smooth functions of $u$ and $\phi(0)=0$.
We first find the local maxima and minima of $|V(u)|^2$; at these points, $\phi$ should be close to, but slightly smaller than, $\left(\frac{1}{2}+2i\right)\pi$ and $\left(\frac{3}{2}+2i\right)\pi$, if $A(u)$ and $B(u)$ monotonically decrease with $u$.
We obtain an initial decomposition by setting $\phi$ to $\left(\frac{1}{2}+2i\right)\pi$ or $\left(\frac{3}{2}+2i\right)\pi$ at the local extrema of $|V(u)|^2$, solving for envelope of $|V(u)|^2$ at these extrema, and then interpolating $A(u)$, $B(u)$ and $\phi(u)$ between the extrema using cubic splines.
We then iteratively refine the decomposition by finding the local extrema of $\left[|V(u)|^2-A(u)\right]/B(u)$ to repeat this process, which corrects the aforementioned bias if one naively sets the phase to $\left(\frac{1}{2}+2i\right)\pi$ or $\left(\frac{3}{2}+2i\right)\pi$ at the local extrema of $|V(u)|^2$, and typically converges within about 5 iterations.

\clearpage

\section{Results for the SANE simulation}
\label{app:SANE}

\begin{figure}[!ht]
    \centering
    \includegraphics[width=0.48\columnwidth]{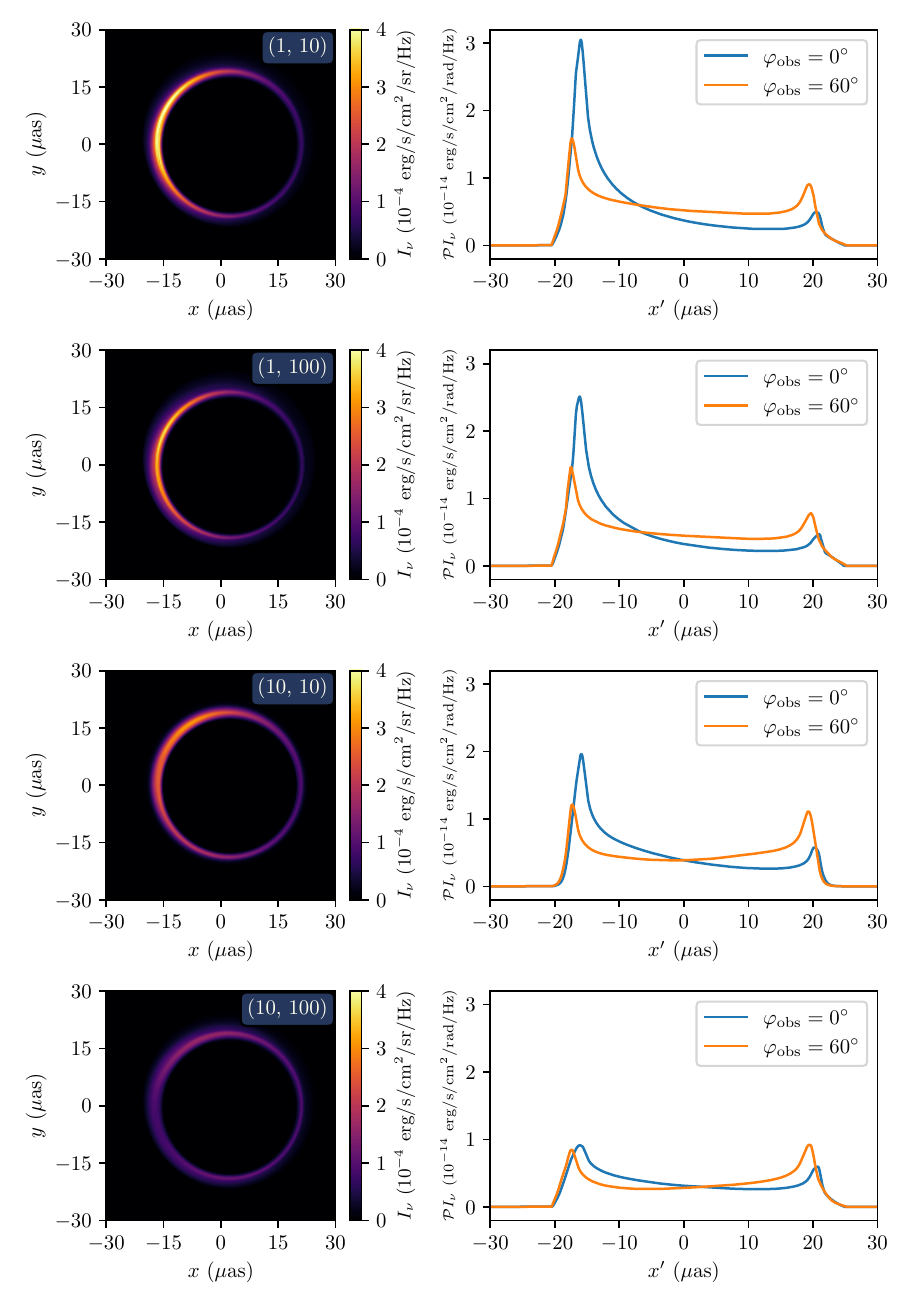}
    \caption{Similar to Fig.~\ref{fig:img-mad}, but for the SANE GRMHD simulation.}
    \label{fig:img-sane}
\end{figure}

We show the results for the SANE simulation in Figs.~\ref{fig:img-sane} and \ref{fig:phase-sane-0}: while the $m=5$ model still consistently outperforms the $m=1$ model for the visibility modeling, they both diverge on shorter baselines than in the MAD simulation with the same electron temperature model.
This is because the $n=1$ subimages generally have ``heavier tails'' (i.e., more intensity well away from the peak of the ring subimage) in SANE models compared with MAD models, as demonstrated in Fig.~\ref{fig:slice-mad-sane}.
In other words, the effective width of the $n=1$ ring is larger in the SANE models, making the perturbative approach break down on smaller baselines.

However, we note that the subimage being non-perturbative does not necessarily mean that the total visibility does not follow our polynomial model, as the approach to decompose the total image into subimages is not unique.
At baselines where the $n=1$ subimage dominates the total visibility, while the $n=1$ subimage and $n\leq 1$ subimage (i.e., the sum of $n=0$ and $n=1$) have the same visibility, the former will be more perturbative than the latter, as the latter also includes the long tail due to the $n=0$ component.
While this $n=0$ tail has negligible contribution to the total visibility on longer baselines, it can still render the total ring non-perturbative since one can no longer expand the trigonometric functions as polynomials in the calculation.
Therefore, throughout this paper, we apply our perturbative method to the $n=1$ subimage, defined via the ${\rm d}z/{\rm d}\lambda=0$ turning points.
As the decomposition is not unique, it may be possible to define an alternative decomposition scheme, with which the $n'=1$ subimage has the same visibility (within some desired baseline window) but even shorter tails than the standard $n=1$ subimage, so that the ring remains perturbative on longer baselines.
Nevertheless, since the standard subimage decomposition seems sufficient for the visibility modeling below $40\,{\rm G}\lambda$ for all the plasma models considered in this work, we leave such an alternative decomposition scheme for future research.

\clearpage

\section{Convergence with respect to the range of radial integration}
\label{app:range}

\begin{figure*}[!ht]
    \centering
    \includegraphics[width=\textwidth]{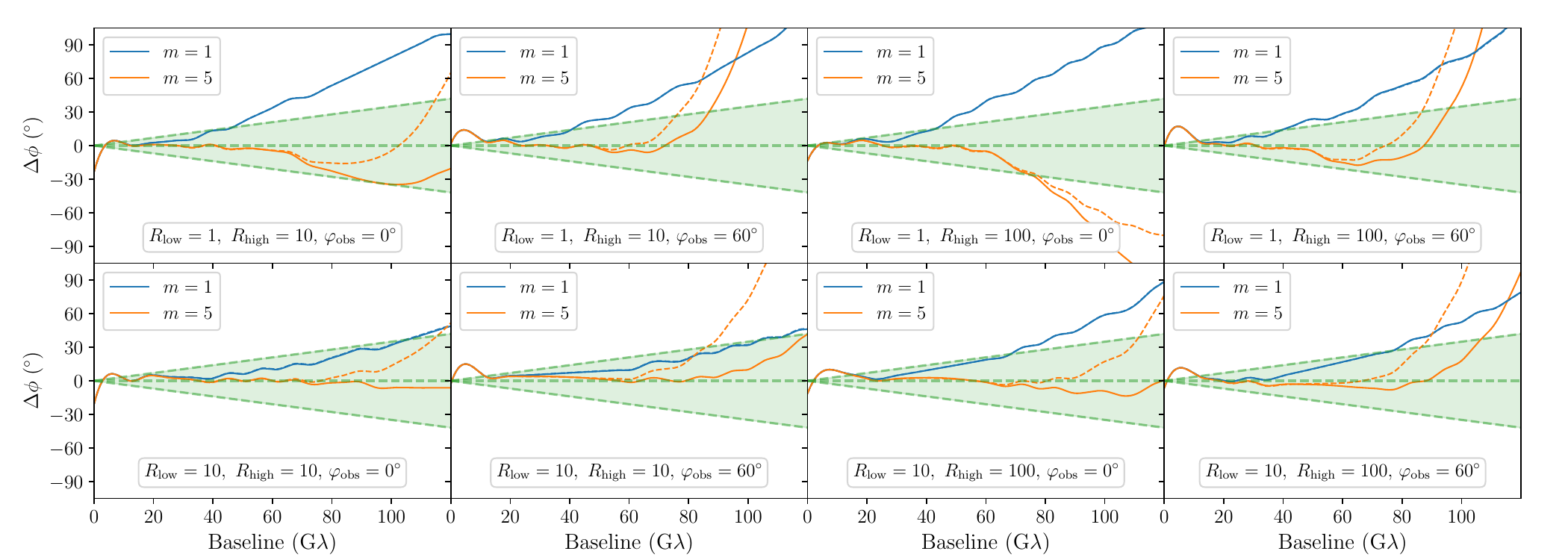}
    \caption{Similar to Fig.~\ref{fig:phase-mad-0}, but comparing the results where the radial integral runs over the whole ring (dashed) versus where it runs between the 0.1\% and 99.9\% radial intensity quantiles (solid).
    We choose to truncate the ring at 0.1\% quantiles on each end, which slightly improves the accuracy of the fifth-order visibility modeling up to longer baselines.}
    \label{fig:phase-mad-2}
\end{figure*}

In Fig.~\ref{fig:phase-mad-2}, we compare the visibility phase calculations with the integral in Eq.~\eqref{eq:Moments} running over $0\leq q\leq 1$ (the whole ring) versus $0.1\%\leq q\leq 99.9\%$; the latter effectively truncates the ring at the 0.1\% radial intensity quantiles on each side.
We find that while the $m=1$ results remains indistinguishable, the $m=5$ results on longer baselines do depend on the range of the radial integral: due to the $(r')^m$ factor in Eq.~\eqref{eq:Moments}, the higher order moments will be more sensitive to the tails of the ring intensity profile.
Throughout this paper, we adopt the $0.1\%\leq q\leq99.9\%$ truncation as our default choice, which slightly improves the accuracy of the fifth order visibility modeling.

\end{document}